\PassOptionsToPackage{table,xcdraw}{xcolor}
\documentclass[12pt,a4paper]{article}
\pdfoutput=1
\usepackage{jheppub}
\usepackage{gensymb}
\DeclareSymbolFont{matha}{OML}{txmi}{m}{it}
\DeclareMathSymbol{\varv}{\mathord}{matha}{118}
\usepackage{amsmath}
\usepackage{amssymb}
\usepackage{graphicx}
\usepackage{subfigure}
\usepackage{pstricks}
\usepackage{bm}
\usepackage{pbox}
\usepackage{placeins}
\usepackage{booktabs}
\usepackage[T1]{fontenc}
\usepackage{footnote}
\usepackage{pdfpages}
\usepackage{hhline}
\usepackage{multirow}
\usepackage{multicol}
\usepackage{enumitem}
\usepackage[toc,page]{appendix}
\allowdisplaybreaks
\usepackage{comment}
\usepackage{float}
\usepackage{slashed}
\usepackage{xcolor}
\usepackage{textcomp}
\usepackage{gensymb}
\usepackage{multirow}
\usepackage[numbers]{natbib}
\usepackage{notoccite}

\usepackage{rotating}
\usepackage{tabularx}
\usepackage[labelfont=bf]{caption} 

\FloatBarrier

\usepackage[many]{tcolorbox}

\renewcommand{\thetable}{\Roman{table}}

\renewcommand{\thetable}{\Roman{table}}
\newcommand{\matrixel}[3]{\left< #1 \vphantom{#2#3} \right|
 #2 \left| #3 \vphantom{#1#2} \right>} 

\definecolor{MyDarkBlue}{rgb}{0.1, 0.1, 0.8} 
\definecolor{MyLightBlue}{rgb}{0.22,0.51,0.9}
\definecolor{MyGreen}{rgb}{0.0, 0.5, 0.0}
\definecolor{BrickRed}{rgb}{0.8, 0.25, 0.33}

\RequirePackage{hyperref}
\hypersetup{colorlinks, citecolor=BrickRed,linkcolor=blue, urlcolor=magenta}

\makeatletter
\gdef\@fpheader{}
\makeatother
\begin{document}

\title{\bf A Parameter Space Exploration of the
Minimal \boldmath{$SU(5)$} Unification}
\author[a,b]{Ilja Dor\v{s}ner,}
\author[c]{Emina D\v{z}aferovi\'c-Ma\v{s}i\'c,}
\author[d]{and Shaikh Saad}

\affiliation[a]{University of Split, Faculty of Electrical Engineering, Mechanical Engineering and\\ Naval Architecture in Split (FESB), Ru\dj era Bo\v{s}kovi\'{c}a 32, HR-21000 Split, Croatia}
\affiliation[b]{CERN, Theoretical Physics Department, CH-1211 Geneva 23, Switzerland}
\affiliation[c]{University of Zagreb, Faculty of Science, Department of Physics, Bijeni\v{c}ka cesta 32, HR-10000 Zagreb, Croatia}
\affiliation[d]{Department of Physics, University of Basel, Klingelbergstr.\ 82, CH-4056 Basel, Switzerland}

\emailAdd{dorsner@fesb.hr, eminadz@gmail.com, shaikh.saad@unibas.ch}
\abstract{
We present phenomenological study of the most minimal realistic $SU(5)$ model that owns its predictivity solely to the gauge symmetry and the representational content. The model is built entirely out of the fields residing in the first five lowest dimensional representations that transform non-trivially under the $SU(5)$ gauge group. It has eighteen real parameters and fourteen phases, all in all, to address experimental observables of the Standard Model fermions and accomplishes that via simultaneous use of three different mass generation mechanisms. Furthermore, it inextricably links the origin of the neutrino mass to the experimentally observed difference between the down-type quark and charged lepton masses. The main predictions of the model are that $(i)$ the neutrinos are Majorana particles, $(ii)$ one neutrino is massless, $(iii)$ the neutrinos have normal mass ordering, and $(iv)$ there are four new scalar multiplets at or below a $120$\,TeV mass scale. A one-loop analysis demonstrates that an improvement of the current $p \rightarrow \pi^0 e^+$ partial lifetime limit by a factor of $2$, $15$, and $96$ would require these four scalar multiplets to reside at or below the $100$\,TeV, $10$\,TeV, and $1$\,TeV mass scales, respectively.
}

\keywords{Grand Unified Theory, Fermion spectrum, Proton decay}


\maketitle

\section{Introduction}
\label{SEC-01}

One of the recurring themes within the elementary particle physics model building community is a quest for simplicity of the proposed scenarios. The premise behind this approach to the model building is that the simpler the scenario is the more predictive and thus testable it becomes. We put to the test this expectation by studying the predictions of the most minimal renormalizable $SU(5)$ model in the literature to date that is still viable~\cite{Dorsner:2019vgf}. We survey the entire parameter space of this model in order to spell out accurate predictions and phenomenological signatures that originate solely from its structure without referral to any additional symmetries and/or assumptions whatsoever.

The main predictions of the model are that $(i)$ the neutrinos are Majorana particles, $(ii)$ one neutrino is massless, $(iii)$ the neutrino mass ordering corresponds to the normal hierarchy, and $(iv)$ there exists a direct link between experimental bound on the proton decay lifetime, as provided by the measurement of the $p \rightarrow \pi^0 e^+$ channel, and the upper bound on the most easily accessible mass scale of new physics. Namely, a one-loop analysis stipulates the existence of four new scalar multiplets at or below a $120$\,TeV mass scale. In fact, an improvement of the current $p \rightarrow \pi^0 e^+$ partial lifetime limit by a factor of $2$, $15$, and $96$ would require these four scalar multiplets to reside below the $100$\,TeV, $10$\,TeV, and $1$\,TeV mass scales, respectively.

Other notable virtues of the model are as follows. The model has eighteen real parameters and fourteen phases, all in all, to address experimentally accessible properties associated with the Standard Model fermions such as masses, mixing angles, CP violating phases, and Majorana phases. It is entirely built out of the first five non-trivial $SU(5)$ representations of the lowest lying dimensionalities. It has only one multiplet that can be identified as the Standard Model Higgs doublet while the proton decay mediating fields are exactly the ones as in the original Georgi-Glashow model~\cite{Georgi:1974sy}.

A simplicity of the model can also be observed from the fact that the neutrino mass matrix is built out of two rank-one matrices whereas the mismatch between the masses of the down-type quarks and charged leptons is given by a single rank-one matrix, where these three matrices have one row matrix in common. It is not only that both the neutrino masses and the observed difference between the masses of the down-type quarks and charged leptons are generated in the most minimal way possible but that they are inextricably linked to each other. This places significant constraints on the model parameters as we discuss in detail later on. It also limits a range of viable values for the CP phase in the neutrino sector.

The manuscript should be seen as a comprehensive extension of the previous analysis of this model~\cite{Dorsner:2019vgf} and is organised as follows. In Section~\ref{SEC-02} we discuss the specifics of the model, such as the particle content and the associated interactions, symmetry breaking effects pertinent to gauge coupling unification, and mass generation mechanisms, to set the stage for the numerical analysis. The procedures behind the numerical study are subsequently discussed and the main results presented in Section~\ref{SEC-03}. We finally conclude in Section~\ref{SEC-04}. 

\section{The model description}
\label{SEC-02}

\subsection{Particle content and notation}
The model is built out of $5_H$, $24_H$, $35_H$, ${\overline{5}_F}_i$, ${10_F}_i$, $15_F$, $\overline{15}_F$, and $24_V$, where subscripts $H$, $F$, and $V$ denote representations comprising scalars, fermions, and gauge bosons, respectively, and $i(=1,2,3)$ is the generation index. This model extends the particle content of the original Georgi-Glashow model~\cite{Georgi:1974sy} with one scalar representation, i.e., $35_H$, and one vector-like fermion representation comprising $15_F$ and $\overline{15}_F$. These two additions overcome three shortcomings of the original Georgi-Glashow model. Namely, these representations $(a)$ generate realistic neutrino masses, $(b)$ create experimentally observed mismatch between the masses of the down-type quarks and charged leptons, and $(c)$ provide viable gauge coupling unification. That is, in a nutshell, the main source of predictivity of this model. 

We summarize the particle content and symbolic notation for the aforementioned irreducible representations and their decompositions under the Standard Model gauge group $SU(3) \times SU(2) \times U(1)$ in Table~\ref{table:fields}. 
\begin{table}
\begin{center}
\begin{tabular}{| c | c | c | c | c | c |}
\hline
$SU(5)$ & Standard Model & $(b_3,b_2,b_1)$ & $SU(5)$ & Standard Model & $(b_3,b_2,b_1)$\\\hline
\hline
 & $\Lambda_1\left(1,2,\frac{1}{2}\right)$ & $\left(0,\frac{1}{6},\frac{1}{10}\right)$ & & $L_i\left(1,2,-\frac{1}{2}\right)$ & $\left(0,1,\frac{3}{5}\right)$\\
\raisebox{2ex}[0pt]{$5_H \equiv \Lambda$} & $\Lambda_3\left(3,1,-\frac{1}{3}\right)$ & $\left(\frac{1}{6},0,\frac{1}{15}\right)$ & \raisebox{2ex}[0pt]{${\overline{5}_F}_i \equiv F_i$} & $d_i^c\left(\overline{3},1,\frac{1}{3}\right)$ & $\left(1,0,\frac{2}{5}\right)$\\ 
\hline
 & $\phi_0 \left(1,1,0\right)$ & $\left(0,0,0\right)$ & & $Q_i\left(3,2,\frac{1}{6}\right)$ & $\left(2,3,\frac{1}{5}\right)$\\
 & $\phi_1 \left(1,3,0\right)$ & $\left(0,\frac{1}{3},0\right)$ & ${10_F}_i \equiv T_i$ & $d_i^c\left(\overline{3},1,-\frac{2}{3}\right)$ & $\left(1,0,\frac{8}{5}\right)$\\
$24_H \equiv \phi$ & $\phi_3\left(3,2,-\frac{5}{6}\right)$ & $\left(\frac{1}{6},\frac{1}{4},\frac{5}{12}\right)$ &  & $e_i^c \left(1,1,1\right)$ & $\left(0,0,\frac{6}{5}\right)$\\\cline{4-6}
 & $\phi_{\overline{3}} \left(\overline{3},2,\frac{5}{6}\right)$ & $\left(\frac{1}{6},\frac{1}{4},\frac{5}{12}\right)$ &  & $\Sigma_1(1,3,1)$ & $\left(0,\frac{4}{3},\frac{6}{5}\right)$\\
  & $\phi_8 \left(8,1,0\right)$ & $\left(\frac{1}{2},0,0\right)$ & $15_F \equiv \Sigma$ & $\Sigma_3\left(3,2,\frac{1}{6}\right)$ & $\left(\frac{2}{3},1,\frac{1}{15}\right)$\\\cline{1-3}
  & $\Phi_1  \left(1,4,-\frac{3}{2}\right)$ & $\left(0,\frac{5}{3},\frac{9}{5}\right)$ & & $\Sigma_6\left(6,1,-\frac{2}{3}\right)$ & $\left(\frac{5}{3},0,\frac{16}{15}\right)$\\\cline{4-6}
  & $\Phi_3  \left(\overline{3},3,-\frac{2}{3}\right)$ & $\left(\frac{1}{2},2,\frac{4}{5}\right)$ & & $\overline{\Sigma}_1\left(1,3,-1\right)$ & $\left(0,\frac{4}{3},\frac{6}{5}\right)$\\
 \raisebox{2ex}[0pt]{$35_H \equiv \Phi$} & $\Phi_6  \left(\overline{6},2,\frac{1}{6}\right)$ & $\left(\frac{5}{3},1,\frac{1}{15}\right)$ & $\overline{15}_F \equiv \overline{\Sigma}$ & $\overline{\Sigma}_3\left(\overline{3},2,-\frac{1}{6}\right)$ & $\left(\frac{2}{3},1,\frac{1}{15}\right)$\\
 & $\Phi_{10}  \left(\overline{10},1,1\right)$ & $\left(\frac{5}{2},0,2\right)$ &  & $\overline{\Sigma}_6\left(\overline{6},1,\frac{2}{3}\right)$ & $\left(\frac{5}{3},0,\frac{16}{15}\right)$\\
\hline  
\end{tabular}
\end{center}
\caption{The field content, $\beta$-function coefficients, and the associated nomenclature at both the $SU(5)$ and the Standard Model levels. $i(=1,2,3)$ is a generation index.}
\label{table:fields}
\end{table}

The symmetry breaking chain is the same as in the original Georgi-Glashow model, i.e., $SU(5) \rightarrow SU(3) \times SU(2) \times U(1) \rightarrow SU(3) \times U(1)_\mathrm{em}$, and the relevant vacuum expectation values are $\langle 24_H \rangle=v_{24}/\sqrt{15} \ \textrm{diag}(1,1,1,-3/2,-3/2)$ and $\langle 5_H \rangle = (0 \quad 0 \quad 0 \quad 0 \quad v_{5})^T$, where $v_5(=174.104$\,GeV) is the Standard Model vacuum expectation value. (The effects associated with vacuum expectation values of the electrically neutral components of $\phi_1$ and $\Phi_1$ scalars are considered to be negligible.)

The Lagrangian of the model, apart from the kinetic terms, is 
\begin{align}
\mathcal{L}\supset &
\left\{+Y^u_{ij}\;T^{\alpha\beta}_iT^{\gamma\delta}_j\Lambda^\rho \epsilon_{\alpha\beta\gamma\delta\rho} 
+Y^d_{ij}\;T^{\alpha\beta}_iF_{\alpha j} \Lambda^{\ast}_\beta 
+Y^{a}_{i}\;\Sigma^{\alpha\beta}F_{\alpha i}  \Lambda^{\ast}_\beta
+Y^{b}_{i}\; \overline{\Sigma}_{\beta\gamma}F_{\alpha i} \Phi^{*\alpha\beta\gamma} \right.
\nonumber \\ &
\left. +Y^{c}_{i}\; T^{\alpha\beta}_i  \overline{\Sigma}_{\beta\gamma}\phi^\gamma_\alpha+\mathrm{h.c.}\right\}
+M_{\Sigma} \overline{\Sigma}_{\alpha\beta} \Sigma^{\alpha\beta}
+y\; \overline{\Sigma}_{\alpha\beta} \Sigma^{\beta\gamma} \phi_\gamma^\alpha \nonumber \\    
&-\mu^2_\Lambda\; \left(\Lambda^*_\alpha\Lambda^\alpha\right) +\lambda^\Lambda_0\;  \left( \Lambda^*_\alpha\Lambda^\alpha  \right)^2 + \mu_1\; \Lambda^*_\alpha\Lambda^\beta \phi^\alpha_\beta +  \lambda^\Lambda_1\;  \left(\Lambda^*_\alpha\Lambda^\alpha\right) \left(\phi^\beta_\gamma\phi_\beta^\gamma\right) + \lambda^\Lambda_2\; \Lambda^*_\alpha\Lambda^\beta \phi_\beta^\gamma\phi_\gamma^\alpha
\nonumber \\ &
-\mu^2_\phi\;  \left(\phi^\beta_\gamma\phi_\beta^\gamma\right)
+\mu_2\; \phi^\alpha_\beta\phi^\beta_\gamma \phi^\gamma_\alpha
+\lambda^\phi_0\; \left( \phi^\beta_\gamma\phi_\beta^\gamma \right)^2
+\lambda^\phi_1\;  \phi^\alpha_\beta\phi^\beta_\gamma \phi^\gamma_\delta\phi^\delta_\alpha 
+\mu^2_\Phi\; \left(\Phi^{*\alpha \beta\gamma}\Phi_{\alpha \beta\gamma}\right)
\nonumber \\ &
+\lambda^\Phi_0\; \left( \Phi^{*\alpha \beta\gamma}\Phi_{\alpha \beta\gamma}\right)^2
+\lambda^\Phi_1\; \Phi^{*\alpha \beta\gamma}\Phi_{\alpha \beta\delta} \Phi^{*\delta\rho\sigma}\Phi_{\rho\sigma\gamma}
+\lambda_0\; \left(\Phi^{*\alpha \beta\gamma}\Phi_{\alpha \beta\gamma}\right) \left(\phi^\delta_\rho\phi_\delta^\rho\right)
\nonumber \\ &
+\lambda_0^\prime\; \left(\Phi^{*\alpha \beta\gamma}\Phi_{\alpha \beta\gamma}\right)  \left(\Lambda^*_\rho\Lambda^\rho\right)
+\lambda_0^{\prime \prime}\; \Phi^{*\alpha \beta\gamma}\Phi_{\beta\gamma\delta}\Lambda^\delta\Lambda^*_\alpha
+\mu_3\; \Phi^{*\alpha \beta\gamma}\Phi_{\beta\gamma\delta}\phi^\delta_\alpha
\nonumber \\ & \label{eq:lagrangian}
+\lambda_1\; \Phi^{*\alpha \beta\gamma}\Phi_{\alpha \delta\rho} \phi^\delta_\beta\phi^\rho_\gamma
+\lambda_2\; \Phi^{*\alpha \beta\rho}\Phi_{\alpha \beta\delta} \phi^\gamma_\rho\phi^\delta_\gamma \;
+ \left\{\lambda^\prime\; \Lambda^\alpha\Lambda^\beta\Lambda^\gamma \Phi_{\alpha\beta\gamma} +\mathrm{h.c.}\right\}\,,
\end{align}
where the first two lines contain two $3\times 3$ matrices, three $1 \times 3$ matrices, and a real number that, together, completely govern the fermion interactions. The relevant matrix elements are denoted with $Y^u_{ij}$, $Y^d_{ij}$, $Y^{a}_{i}$, $Y^{b}_{i}$, $Y^{c}_{i}$, and $y$, where $i,j=1,2,3$. The Greek alphabet indices $\alpha, \beta, \gamma, \delta,\rho, \sigma=1,\ldots,5$ stand for the $SU(5)$ contractions. 

It is possible to freely rotate $SU(5)$ representations, prior to the breaking of $SU(5)$  down to the Standard Model gauge group, in order to choose suitable basis for the parameter counting. It is thus convenient to simultaneously redefine ${\overline{5}_F}_i$ and ${10_F}_i$ in such a way as to render $Y^d$ in the second contraction of the first line of Eq.~\eqref{eq:lagrangian} diagonal and real. (In fact, the entries in $Y^d$ will represent the actual Yukawa couplings of the charged leptons, as we describe in Section~\ref{charged_masses}.) It is also convenient to rotate $15_F$ to remove one phase in a complex matrix $Y^{c}$. We choose this phase to be the one in the $Y^{c}_{3}$ element. This exhausts available redefinitions of the $SU(5)$ representations. 

The up-type quark mass matrix is proportional to the symmetric combination given by $Y^u+Y^{u\,T}$, whereas the neutrino mass matrix is proportional to the sum of two rank-one matrices with elements $Y^{a}_{i} Y^{b}_{j}$ and $Y^{b}_{i} Y^{a}_{j}$, respectively, where $Y^{a}$ and $Y^{b}$ are complex row matrices. The down-type quark mass matrix, on the other hand, is a linear combination of the diagonal matrix $Y^d$ and one rank-one matrix with elements $Y^{c}_{i} Y^{a}_{j}$. Finally, $y$ is simply a real number and since it contributes only to the mass splitting of the vector-like fermions in $15_F$ and $\overline{15}_F$, we do not count it as the relevant parameter towards the Standard Model fermion mass input. There are thus eighteen real parameters and fourteen phases available, in the Yukawa sector of the model, to address experimentally accessible properties of the Standard Model fermions. 

As we will expand upon later, the up-type quarks and charged leptons get the mass purely from the usual Higgs mechanism, neutrinos get the mass at the one-loop level through an exchange of the vector-like fermions comprising $15_F$ and $\overline{15}_F$, whereas the down-type quarks get the masses from the Higgs mechanism and the interactions with the aforementioned vector-like fermions. 

We will now briefly outline the most essential symmetry breaking effects one needs to take into account when discussing the gauge coupling unification within this model.

\subsection{Symmetry breaking and unification} 
\label{GCU}

The relevant degrees of freedom that are {\it a priori} not known and that can enter the gauge coupling unification analysis within our model are the masses of $\Phi_1, \Phi_3, \Phi_6, \Phi_{10} \in 35_H$, $\Sigma_1, \Sigma_3, \Sigma_6 \in 15_F$, $\phi_1, \phi_8 \in 24_H$, and $\Lambda_3 \in 5_H$. Scalar fields $\phi_3$ and $\phi_{\overline{3}}$ from $24_H$, on the other hand, provide necessary degrees of freedom to the proton mediating gauge bosons in $24_V$, during the $SU(5)$ symmetry breaking process, and are thus only indirectly featured in the unification study. (See Table~\ref{table:fields} for the relevant notation and the field transformation properties.) 

There are, however, two particular mass relations induced by the $SU(5)$ symmetry breaking that need to be satisfied within this model. The first one concerns three Standard Model vector-like fermion multiplets comprising $15_F$ and $\overline{15}_F$. Namely, it is the last two terms in the second line of Eq.~\eqref{eq:lagrangian} that generate mass contributions for these three fermion multiplets thus creating one mass relation that reads
\begin{equation}
\label{eq:mass_relation_a}
M_{\Sigma_6}=2 M_{\Sigma_3}-M_{\Sigma_1}\,.
\end{equation}
Analogously, there are only three linearly independent contractions in Eq.~\eqref{eq:lagrangian} that generate masses for four multiplets in $35_H$. This yields the second mass relation of the form
\begin{equation}
\label{eq:mass_relation_b}
M^2_{\Phi_{10}}=M^2_{\Phi_1}-3M^2_{\Phi_3}+3 M^2_{\Phi_6}\,.
\end{equation}

The gauge coupling unification analysis, as we demonstrate later in Section~\ref{unification}, requires $\Phi_1$ to be very heavy, i.e., $M_{\Phi_1} \gg v_5$, whereas ${\Phi_3}$ and ${\Phi_6}$ prefer to be light if the unification scale $M_\mathrm{GUT}$ is to be sufficiently large in view of the proton decay constraints. This simply means, through the use of Eq.~\eqref{eq:mass_relation_b}, that $\Phi_1$ and $\Phi_{10}$ are consequently heavy, and of the same mass, whereas the fields $\Phi_3$ and $\Phi_6$ are mass degenerate and light within a viable parameter space. In other words, there are only two mass scales associated with the fields residing in $35_H$. Moreover, in the regime of interest, i.e., when the model can accommodate neutrino masses and provide gauge coupling unification, the fields $\Sigma_1$, $\Sigma_6$, and $\Sigma_3$ tend to be mass degenerate with $M_{\Sigma_1},M_{\Sigma_6},M_{\Sigma_3} \gg v_5$. This, again, means that there is only one mass scale associated with the vector-like fermions that reside in $15_F$ and $\overline{15}_F$. This common scale for the full $SU(5)$ multiplet will not affect the value of the unification scale $M_\mathrm{GUT}$ but will leave an imprint on the value of the $SU(5)$ gauge coupling constant $\alpha_\mathrm{GUT}$ instead. After we include $\phi_1$, $\phi_8$, and $\Lambda_3$ into this parameter counting there are effectively only six mass scales that govern the gauge coupling unification in this model. (One also needs to insure that a proton does not decay too rapidly. To that end, we observe that one needs to have $M_{\Lambda_3} \geq 3 \times 10^{11}$\,GeV in order for the scalar induced proton decay to be under control~\cite{Dorsner:2012uz}.) We will later show that if one aims to find the largest possible value of $M_\mathrm{GUT}$ one effectively ends up with only three relevant mass scales. We finally note that the masses of the proton decay mediating gauge bosons in $24_V$ are equal to the unification scale $M_\mathrm{GUT}$, where 
\begin{equation}
M^2_\mathrm{GUT}=\frac{5 \pi}{6} \alpha_\mathrm{GUT} v^2_{24}.
\end{equation}

With this in mind, we turn our attention to the issue of the fermion mass generation.

\subsection{Neutrino mass generation}
\label{neutral_masses}

\begin{figure}[t!]
\centering
\includegraphics[width=0.47\textwidth]{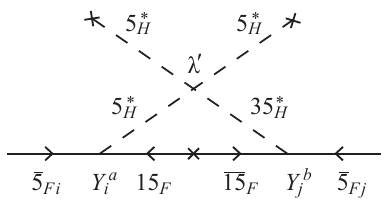}
\includegraphics[width=0.47\textwidth]{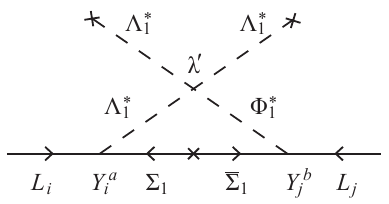}
\caption{The Feynman diagrams of the leading order contribution towards Majorana neutrino masses at the $SU(5)$ (left panel) and the Standard Model (right panel) levels.}
\label{fig:diagram}
\end{figure}

The neutrino masses, in our model, are of the Majorana nature. The leading order contribution is generated at the one-loop level via the $d=5$ operator~\cite{Babu:2009aq,Bambhaniya:2013yca}. The relevant Feynman diagrams, both at the $SU(5)$ and the Standard Model levels, are shown in Fig.~\ref{fig:diagram}. 

The neutrino mass matrix $M_N$, in the regime of interest when $M_{\Sigma_1}, M_{\Phi_1} \gg v_5$, reads 
\begin{equation}
(M_N)_{ij}\approx \frac{\lambda^{\prime}v_5^2}{8 \pi^2} (Y^a_iY^b_j+Y^b_iY^a_j) \frac{M_{\Sigma_1}}{M^2_{\Sigma_1}-M^2_{\Phi_1}}
\ln \left( \frac{M^2_{\Sigma_1}}{M^2_{\Phi_1}} \right)  = m_0 (Y^a_iY^b_j+Y^b_iY^a_j)\,. \label{eq:massnu}
\end{equation}
Clearly, $M_N$ is constructed out of two rank-one matrices with elements $Y^a_iY^b_j$ and $Y^b_iY^a_j$ in the most minimal way imaginable. Moreover, all additional contributions towards neutrino mass matrix, although heavily suppressed and thus completely irrelevant, are also proportional to the same combination of the Yukawa couplings. These facts guarantee with certainty that one of the neutrinos is a massless particle. 

A viable explanation of the neutrino mass scale roughly requires that $m_0 \geq \sqrt{\Delta m^2_{31}}/2$, where $\Delta m^2_{31}$ is the largest of the two neutrino mass squared differences, as measured in the neutrino oscillation experiments. This requirement places an additional constraint on the parameter space spanned by $M_{\Phi_1}$ and $M_{\Sigma_1}$ that will be explored later on in Section~\ref{SEC-03}. In fact, we can be even more accurate in assessing the available parameter space to address the neutrino masses in this model. We describe this procedure only for the normal ordering of the neutrino masses in what follows. 

The neutrino mass matrix elements, in this model, are
\begin{equation}
(M_N)_{ij}
= m_0 \left(
Y^a_i Y^b_j+Y^b_i Y^a_j
\right)= (N\;
 \mathrm{diag}(0,m_2,m_3)\;
N^T)_{ij}\,, 
\label{eq:nu}
\end{equation}
where $m_2$ and $m_3$ are neutrino mass eigenstates and $N$ is a unitary matrix. Since we work in the basis where the charged leptons are already in the mass eigenstate basis we can write $N$ as
\begin{equation}
N=\begin{pmatrix} e^{i\gamma_1}&0&\\0&e^{i\gamma_2}&0\\0&0&e^{i\gamma_3} \end{pmatrix} V_\mathrm{PMNS}^*,
\end{equation}
where $V_\mathrm{PMNS}$ is the Pontecorvo-Maki-Nakagawa-Sakata (PMNS) unitary mixing matrix with three mixing angles, one CP violating Dirac phase, and two Majorana phases. One can invert Eq.~\eqref{eq:nu} using results of Refs.~\cite{Cordero-Carrion:2018xre,Cordero-Carrion:2019qtu} to obtain appropriate forms of $Y^a$ and $Y^b$. Namely, the normal ordering yields 
\begin{equation}
\label{eq:perturbativity}
Y^{a\,T}=\frac{1}{\sqrt{2}} \begin{pmatrix}
i\;r_2\;N_{12}+r_3\;N_{13}\\
i\;r_2\;N_{22}+r_3\;N_{23}\\
i\;r_2\;N_{32}+r_3\;N_{33}
\end{pmatrix},\;\;
Y^{b\,T}=\frac{1}{\sqrt{2}} \begin{pmatrix}
-i\;r_2\;N_{12}+r_3\;N_{13}\\
-i\;r_2\;N_{22}+r_3\;N_{23}\\
-i\;r_2\;N_{32}+r_3\;N_{33}
\end{pmatrix}\;,
\end{equation}
where $r_2=\sqrt{m_2/m_0}$ and $r_3=\sqrt{m_3/m_0}$. There are currently six phases in Eq.~\eqref{eq:perturbativity} that one can freely vary for the given $M_{\Phi_1}$, $M_{\Sigma_1}$, and $\lambda^{\prime}$ to check the perturbativity of the Yukawa coupling elements in $Y^a$ and $Y^b$. These phases are $\gamma_1$, $\gamma_2$, $\gamma_3$, one CP violating phase ($\delta^\mathrm{PMNS}$), and two Majorana phases in $V_\mathrm{PMNS}$. We defer the outcome of this analysis to Section~\ref{SEC-03}. Here we only note that the fact that there are six arbitrary phases in Eq.~\eqref{eq:perturbativity} is expected since the six real parameters in $Y^a$ and $Y^b$ have been traded for three PMNS angles and three neutrino masses during the inversion procedure. 

\subsection{Charged fermion masses}
\label{charged_masses}

A presence of the vector-like fermions comprising $15_F$ and $\overline{15}_F$ induces experimentally observed mismatch between the masses of the charged leptons and the down-type quarks. The mismatch itself is due to the physical mixing between the vector-like fermions and fermions in ${10_{F}}_i$. (The effect of this type of mixing on the charged fermion masses has been studied in Ref.~\cite{Oshimo:2009ia} within the context of a supersymmetric $SU(5)$ framework.) Namely, since the quark doublets $Q_i$ in ${10_F}_i$ and $\Sigma_3$ in $15_F$ transform in the same way under the Standard Model gauge group, as can be seen from Table~\ref{table:fields}, these states interact at the $SU(5)$ symmetry breaking level, where the relevant mixing term explicitly reads 
\begin{align}
\mathcal{L} \supset \frac{1}{4}\sqrt{\frac{10}{3}}v_{24} Y^c_i\; Q_i\overline{\Sigma}_3\,.
\end{align}

The electroweak symmetry breaking induces additional mixing terms between the vector-like fermions and fermions in ${\overline{5}_{F}}_i$ and ${10_{F}}_i$ whenever these fermions transform in the exact same way under $SU(3) \times U(1)_\mathrm{em}$, where the induced terms are all proportional to $v_5$. The relevant decomposition under the $SU(3) \times U(1)_\mathrm{em}$ gauge group is $Q_i=u_i(3,2/3)+d_i(3,-1/3)$, $L_i=e_i(1,-1)+\nu_i(1,0)$, $\Sigma_3=\Sigma^u(3,2/3)+\Sigma^d(3,-1/3)$, and $\Sigma_1=\Sigma^{\nu}(1,0)+\Sigma^{e^c}(1,1)+ \Sigma^{e^ce^c}(1,2)$, where the second number in the parentheses represents electric charge in units of absolute value of the electron charge. The aforementioned symmetry breaking effects thus yield the following mass terms for the charged fermions
\begin{align}
\mathcal{L} \supset &
\begin{pmatrix}u_i&\Sigma^u\end{pmatrix}
\begin{pmatrix}
4 v_5 (Y^{u}_{ij}+Y^{u}_{ji}) & \frac{1}{4}\sqrt{\frac{10}{3}}v_{24} Y^c_i \\
0 &M_{\Sigma_3}
\end{pmatrix}
\begin{pmatrix}u^c_j\\\overline{\Sigma}^u\end{pmatrix} 
\label{MASS-FORMULA}  \\& 
+\begin{pmatrix}d_i&\Sigma^d\end{pmatrix}
\begin{pmatrix}
v_5 Y^d_{ij} & \frac{1}{4}\sqrt{\frac{10}{3}}v_{24} Y^c_i\\
v_5 Y^a_j & M_{\Sigma_3}
\end{pmatrix}
\begin{pmatrix}d^c_j\\\overline{\Sigma}^d\end{pmatrix}
+\begin{pmatrix}e_i&\overline{\Sigma}^{e^c}\end{pmatrix}
\begin{pmatrix}
v_5 Y^d_{ji} & v_5 Y^a_i\\ 0 &M_{\Sigma_1}
\end{pmatrix}
\begin{pmatrix}e^c_j\\\Sigma^{e^c}\end{pmatrix}\,. \nonumber 
\end{align}

The gauge coupling unification considerations, coupled with the need to generate correct neutrino mass scale, require states $\Sigma^{u,d,e^c}$ to be very heavy and we can safely integrate them out. We accordingly find, in the limit when $v_{24} Y^c, M_{\Sigma_1}, M_{\Sigma_3} \gg v_5$, that the mass matrices for the up-type quarks ($M_U$), down-type quarks ($M_D$), and charged leptons ($M_E$) are 
\begin{align}
&M_U=\left( \mathbb{I}+\delta^{\prime 2}\;Y^c{Y^c}^{\dagger} \right)^{-\frac{1}{2}} 4 v_5 (Y^{u}+Y^{u\,T}),  \label{massu}
\\
&M_D=\left( \mathbb{I}+\delta^{\prime 2}\;Y^c{Y^c}^{\dagger} \right)^{-\frac{1}{2}} v_5 \left( Y^d + \delta^\prime\; Y^cY^a  \right),  \label{massd}
\\
&M_E=v_5 Y^{d\,T} \label{masse},
\end{align}
where $\delta^\prime \equiv \sqrt{10/3} v_{24}/(4 M_{\Sigma_3})$ is a dimensionless parameter and $\mathbb{I}$ is the $3\times 3$ identity matrix. It turns out that the contributions proportional to $\delta^{\prime 2}\;Y^c{Y^c}^{\dagger}$ are completely negligible in the parameter space of interest. This allows us to write that 
\begin{align}
&M_U= 4 v_5 (Y^{u}+Y^{u\,T})\,,  \label{eq:massu}
\\
&M_D= v_5 \left( Y^d + \delta^\prime\; Y^cY^a  \right)\,,  \label{eq:massd}
\\
&M_E=v_5 Y^{d\,T}\,, \label{eq:masse}
\end{align}
while the masses of the heavy vector-like fermions are
\begin{equation}
\label{eq:heavy}
M_{\Sigma^u}=M_{\Sigma^d}=M_{\Sigma_3}\left( 1+\delta^{\prime 2} \; {Y^c}^{\dagger}Y^c \right)^{\frac{1}{2}}\approx M_{\Sigma_3},\;
M_{\Sigma^{e^c}}=M_{\Sigma^{e^ce^c}}=M_{\Sigma^{\nu}}= M_{\Sigma_1}.
\end{equation}
Note that the masses of vector-like fermions are not affected by the interaction with the Standard Model fermions, thus preserving the mass relation of Eq.~\eqref{eq:mass_relation_a}. 

To summarize, the model uses one vector-like set of fermions in $15_F$ and $\overline{15}_F$ together with $35_H$ to simultaneously $(a)$ generate neutrino masses, $(b)$ create viable mismatch between the down-type quark and charged lepton masses, and $(c)$ provide gauge coupling unification. 

We are finally in a position to discuss the numerical analysis of the model in view of these requirements. 

\section{Numerical analysis}
\label{SEC-03}

Our numerical exploration of the entire parameter space of the model comprises three distinct steps. We briefly outline each of these steps in what follows before we provide an in-depth description in subsequent sections. 

We first look at a viable gauge coupling unification at the one-loop level. To that end, we freely vary the masses of $\Phi_1, \Phi_3, \Phi_6, \Phi_{10} \in 35_H$, $\Sigma_1, \Sigma_3, \Sigma_6 \in 15_F$, $\phi_1, \phi_8 \in 24_H$, and $\Lambda_3 \in 5_H$, while taking into account additional constraints discussed in Section~\ref{GCU}, to find the largest possible value of $M_\mathrm{GUT}$. This approach gives the most conservative representation of the available parameter space since the largest possible unification scale corresponds to the largest possible nucleon lifetimes one would need to probe to test the model. We always set a lower limit on the mass(es) of the new physics state(s), before we numerically look for the viable unification points, to explore the possible connection between the most accessible scale of new physics and $M_\mathrm{GUT}$. To that end, we introduce a mass parameter $M \equiv \min(M_J)$, where $J=\Phi_1, \Phi_3, \Phi_6, \Phi_{10}, \Sigma_1, \Sigma_3, \Sigma_6, \phi_1, \phi_8, \Lambda_3$, and present our findings when $M \geq 1$\,TeV, $M \geq 10$\,TeV, and $M \geq 100$\,TeV. It is already at this stage that the part of potentially viable parameter space can be discarded. Namely, since the neutrino mass scale explicitly depends on $M_{\Phi_1}$ and $M_{\Sigma_1}$ via $m_0$ parameter of Eq.~\eqref{eq:massnu} it is easy to construct a two-dimensional parameter space spanned by $M_{\Phi_1}$ and $M_{\Sigma_1}$ where one could, at least in principle, hope for realistic explanation of neutrino masses, with perturbative couplings, within this model.

Once we find all the unification points that also allow for generation of viable neutrino mass scale we implement the second step of our numerical analysis. Namely, we run the masses and mixing parameters of the Standard Model charged fermions to $M_\mathrm{GUT}$ using the factual new physics mass spectrum associated with a given unification point to account for all the threshold corrections between the low scale and $M_\mathrm{GUT}$ and then perform an accurate numerical fit of the Standard Model observables. (We do not run the neutrino observables due to the fact that the running effect is not very significant and use the associated low-energy input for the fitting procedure instead.) The charged fermion mass renormalization group running is performed at the one-loop level~\cite{Arason:1991ic}. We note that one can separate the gauge coupling unification study from the running of the Standard Model charged fermion parameters, at this level of accuracy, since the latter provides feedback to the former only at the two-loop level whereas the former impacts the latter already at the one-loop level.

The third step of our analysis begins upon completion of the numerical fit of the Standard Model fermion observables for all viable unification points. Namely, we look into constraints due to the proton decay signatures for every single point that corresponds to realistic gauge coupling unification and viable description of the Standard Model fermion observables. This allows us to produce accurate constraint since we have, at our disposal, all the relevant input parameters for such an analysis including $M_\mathrm{GUT}$, $\alpha_\mathrm{GUT}$, unitary transformations of the Standard Model fermions, Yukawa couplings, short-distance coefficients, et cetera. We find that the most stringent experimental limit, i.e., the limit on the $p \rightarrow \pi^0 e^+$ partial lifetime, provides the best constraint on the available parameter space through the predictions for the gauge boson mediated proton decay. 

\begin{figure}[th!]
\centering
\includegraphics[width=0.53\textwidth]{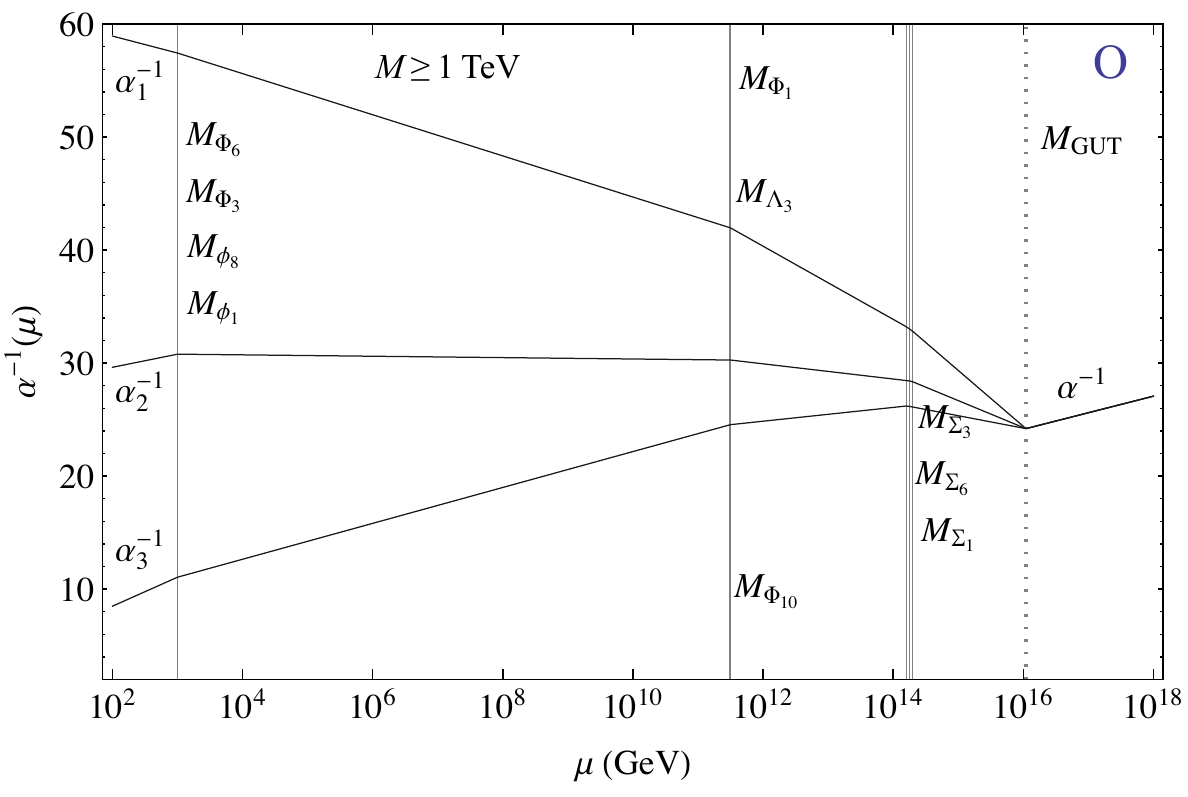}
\includegraphics[width=0.45\textwidth]{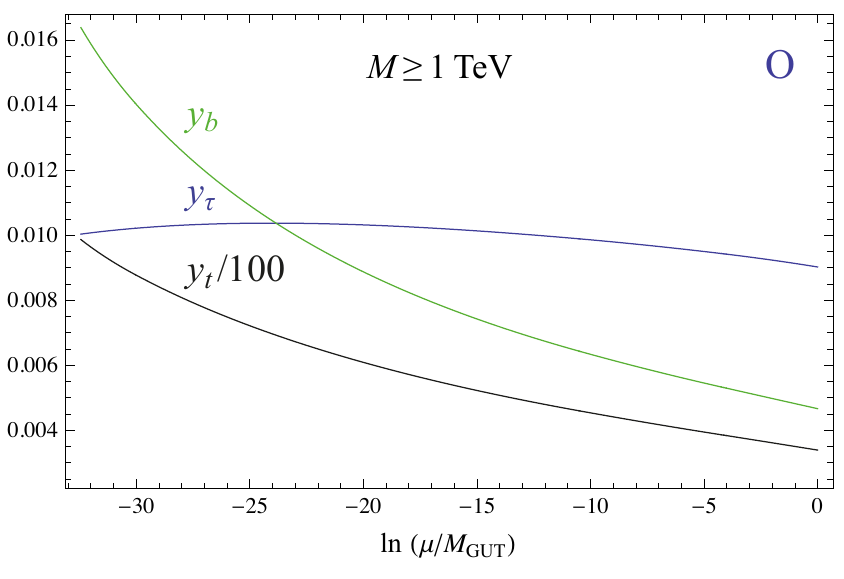}
\includegraphics[width=0.53\textwidth]{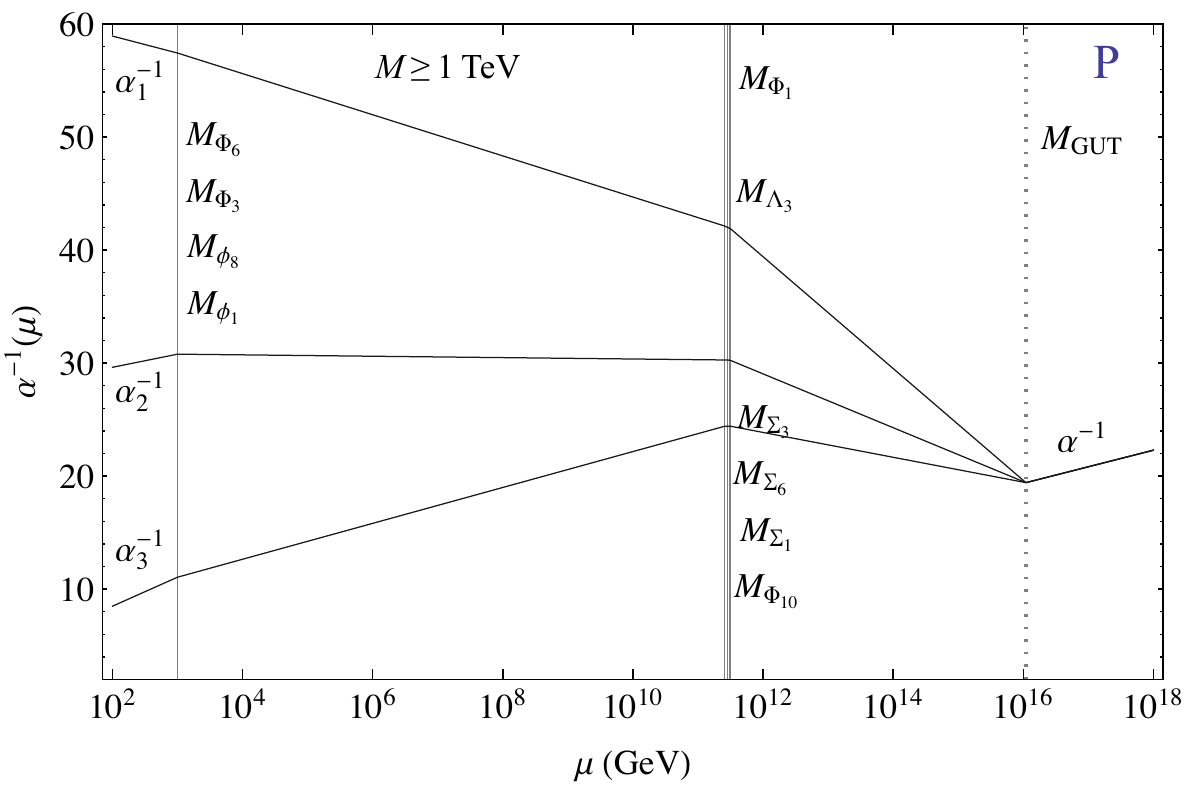}
\includegraphics[width=0.45\textwidth]{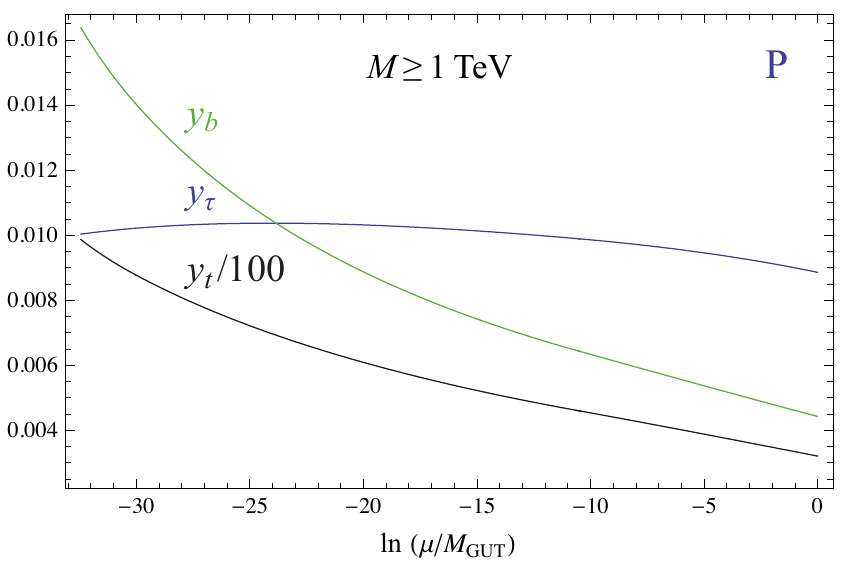}
\includegraphics[width=0.53\textwidth]{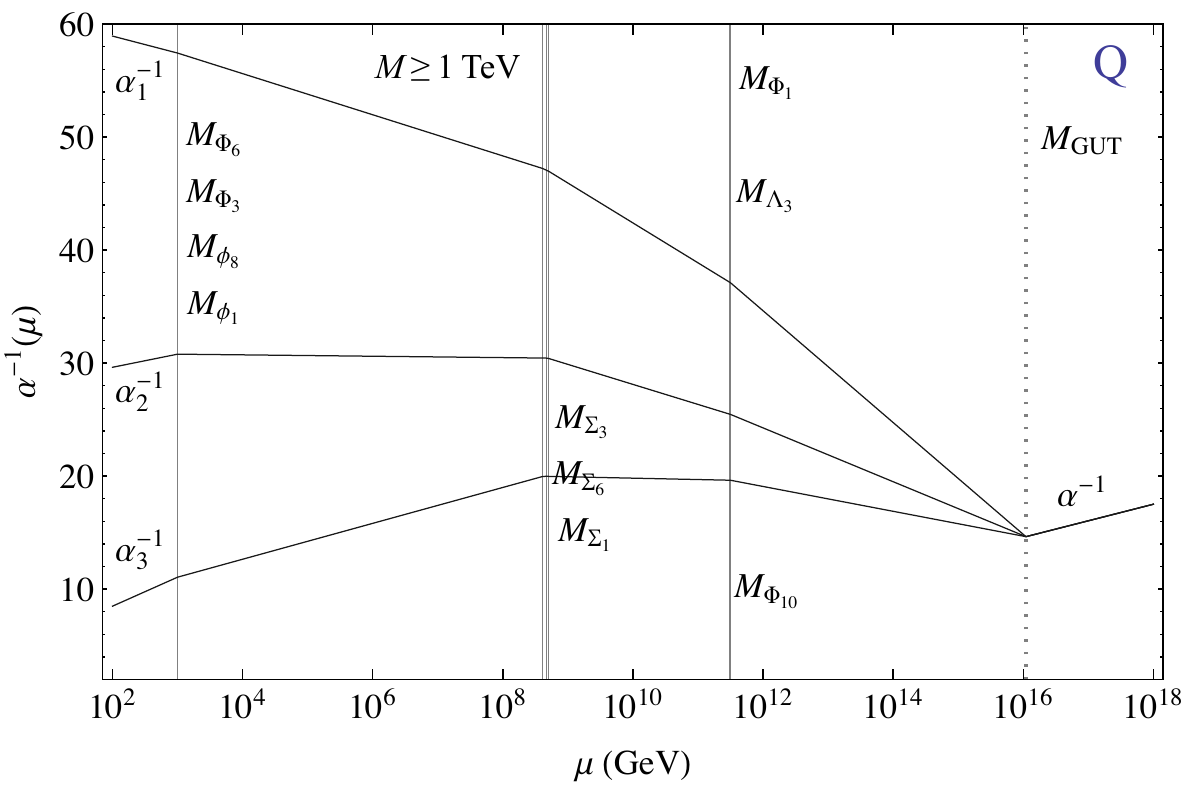}
\includegraphics[width=0.45\textwidth]{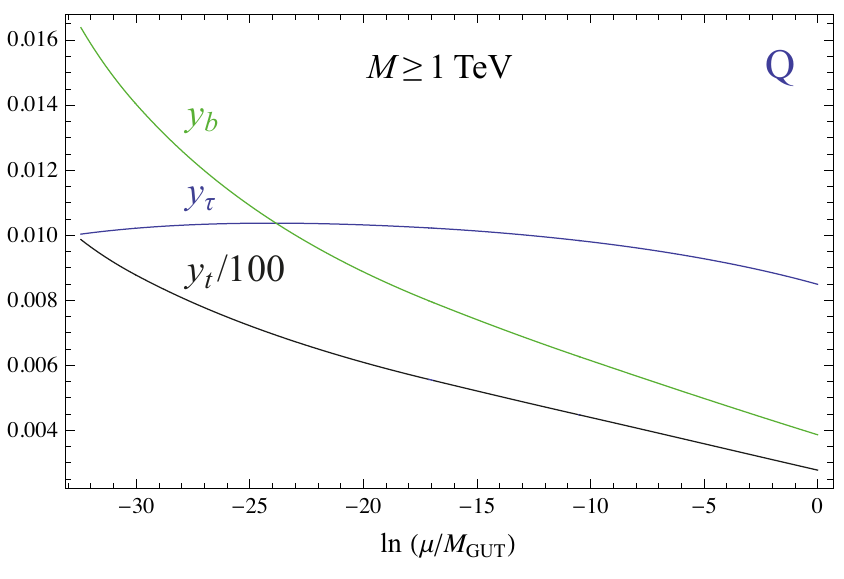}
\caption{The gauge coupling unification for points O, P, and Q (left panels) and the associated renormalization group running of the $\tau$ $(y_\tau)$, $b$ $(y_b)$, and $t$  $(y_t)$ Yukawa couplings (right panels). For additional details see the text.}
\label{fig:OPQ}
\end{figure}

What we are left with, at the end of these three steps, is a viable set of unification points that is in agreement with all currently accessible experimental results and that is what we present in left panels of Fig.~\ref{fig:OPQ}. (We explain the details related to Fig.~\ref{fig:OPQ} in Sections~\ref{unification}, \ref{mass_fit}, \ref{proton_decay}, and~\ref{results}.) It is important to note that any further improvement in experimental determination of the Standard Model parameters, such as the actual determination of the neutrino masses, measurement of the CP phase in the leptonic sector, or an input on Majorana phases, will add to the precision of the model's predictions.

\subsection{Unification analysis}
\label{unification}

To find the unification scale $M_{\mathrm{GUT}}$ of the Standard Model gauge couplings $\alpha_1$, $\alpha_2$, and $\alpha_3$, associated with $U(1)$, $SU(2)$, and $SU(3)$, respectively, and the mass spectrum of the $SU(5)$ model for a corresponding unification point we proceed as follows. We first define coefficients $B_{ij}$ through $B_{ij}=\sum_{J} (b^J_{i}-b^J_{j}) r_{J}$, where $b^J_{i}$ are the $\beta$-function coefficients of a particle $J$ with mass $M_J$, $r_J=\ln
(M_{\mathrm{GUT}}/M_{J})/ \ln (M_{\mathrm{GUT}}/M_{Z})$, and $J=\Phi_1, \Phi_3, \Phi_6, \Phi_{10}, \Sigma_1, \Sigma_3, \Sigma_6, \phi_1, \phi_8, \Lambda_3$. (The relevant $\beta$-function coefficients $b_i$, where $i=1,2,3$, are given in Table~\ref{table:fields}.) We then simultaneously solve the following two equations~\cite{Giveon:1991zm}
\begin{align}
\label{eq:a}
\frac{B_{23}}{B_{12}}&=\frac{5}{8}
\frac{\sin^2
\theta_W-\alpha(M_Z)/\alpha_S(M_Z)}{3/8-\sin^2 \theta_W}\,,\\
\label{eq:b}
\ln \frac{M_{\mathrm{GUT}}}{M_Z}&=\frac{16 \pi}{5
\alpha(M_Z)} \frac{3/8-\sin^2 \theta_W}{B_{12}}\,.
\end{align}
To that end, we freely vary the masses of $\Phi_1, \Phi_3, \Phi_6, \Phi_{10} \in 35_H$, $\Sigma_1, \Sigma_3, \Sigma_6 \in 15_F$, $\phi_1, \phi_8 \in 24_H$, and $\Lambda_3 \in 5_H$, while taking into account additional constraints discussed in Section~\ref{GCU}, to find the largest possible value of $M_\mathrm{GUT}$, where we use  $M_Z=91.1876$\,GeV, $\alpha_S(M_Z)=0.1193\pm0.0016$, $\alpha^{-1}(M_Z)=127.906\pm0.019$, and
$\sin^2 \theta_W=0.23126\pm0.00005$ as our input parameters~\cite{Agashe:2014kda}. 

\begin{figure}[th!]
\centering
\includegraphics[width=0.48\textwidth]{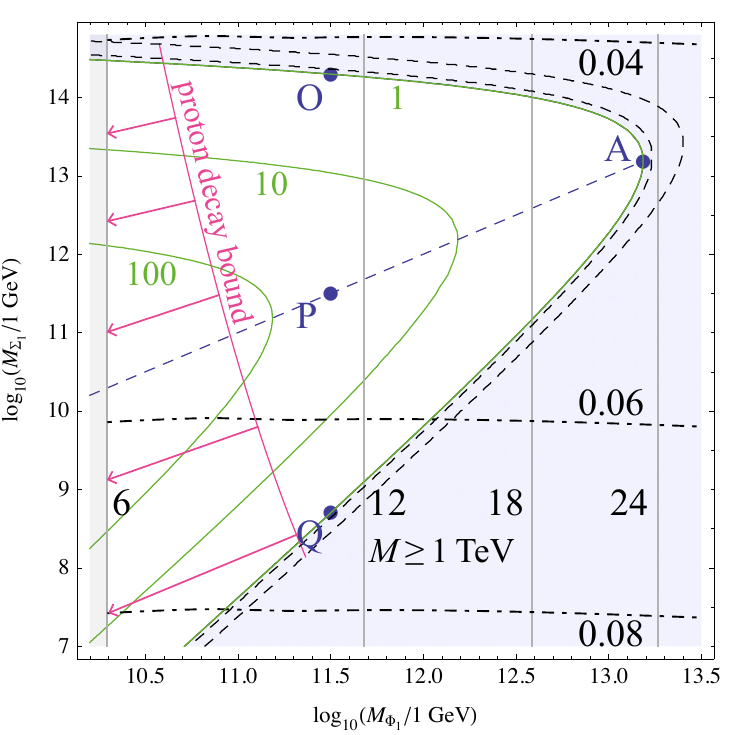}
\includegraphics[width=0.5\textwidth]{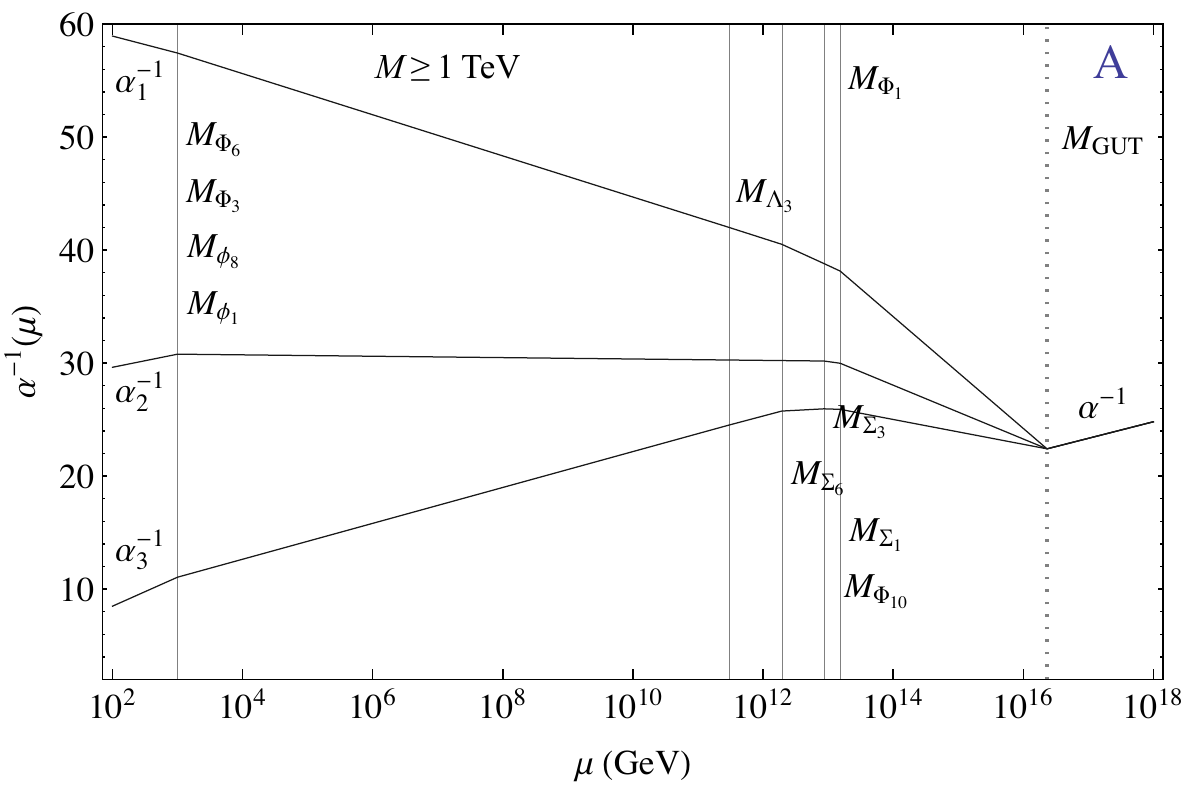}
\includegraphics[width=0.48\textwidth]{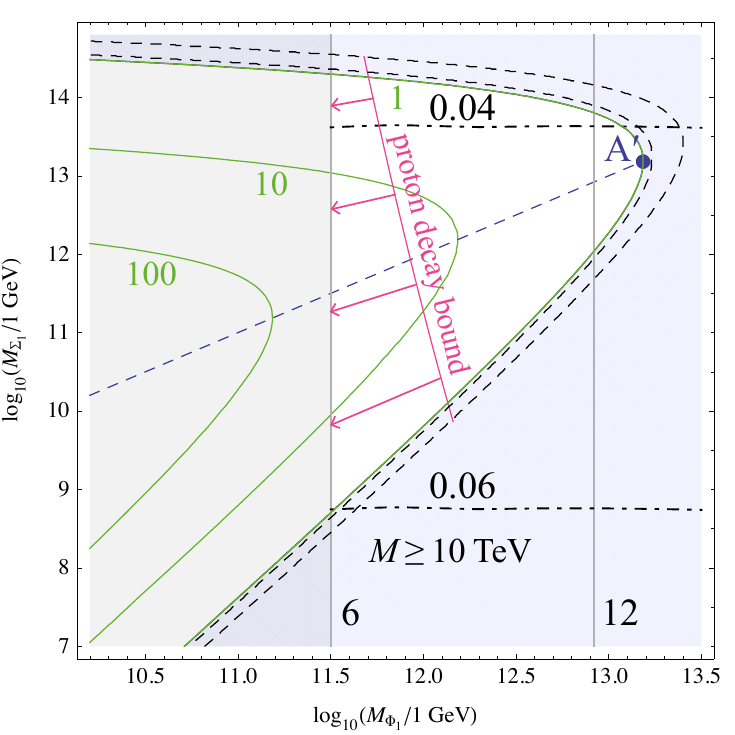}
\includegraphics[width=0.5\textwidth]{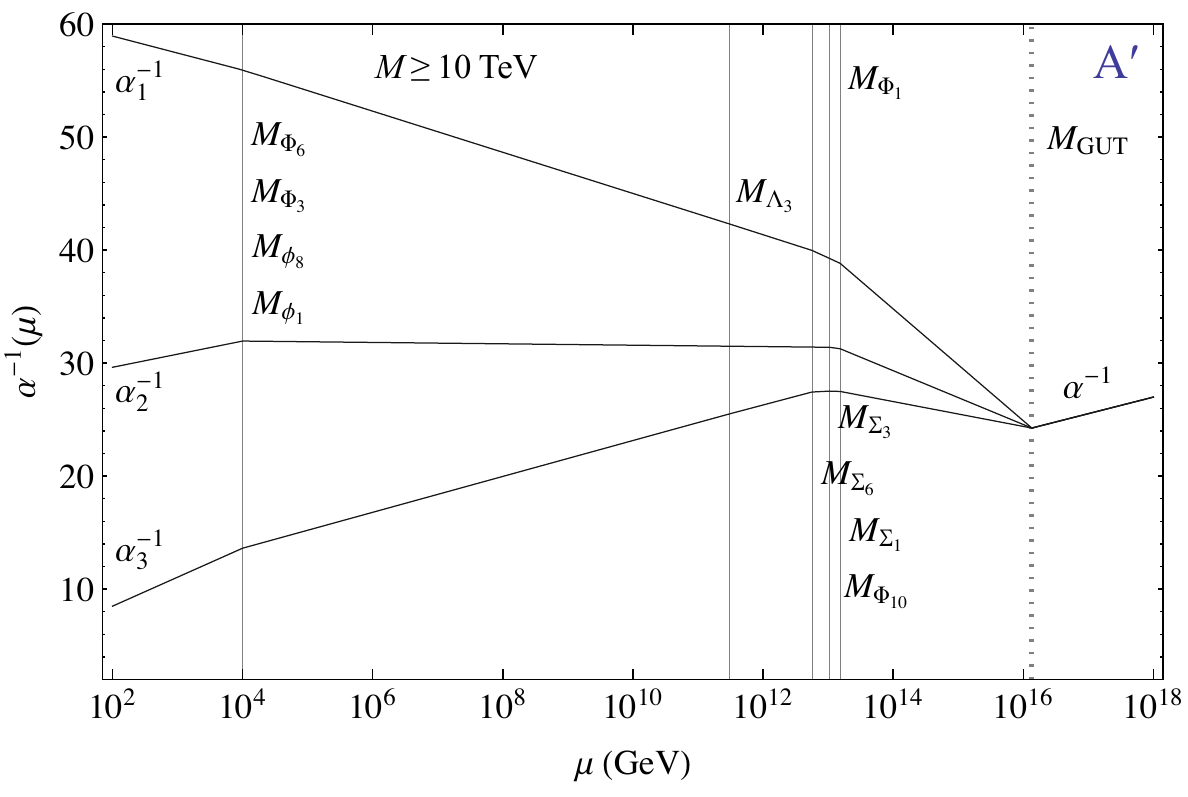}
\includegraphics[width=0.48\textwidth]{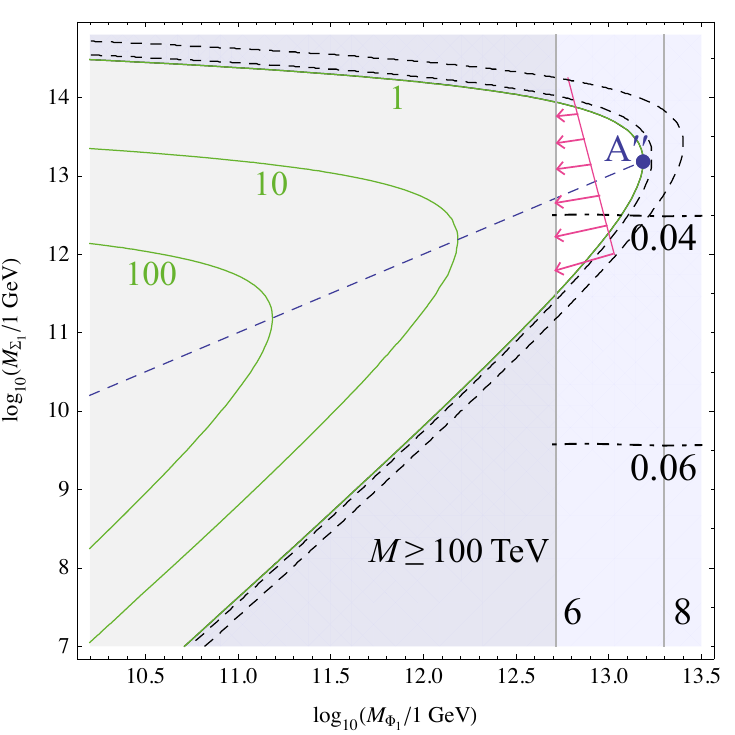}
\includegraphics[width=0.5\textwidth]{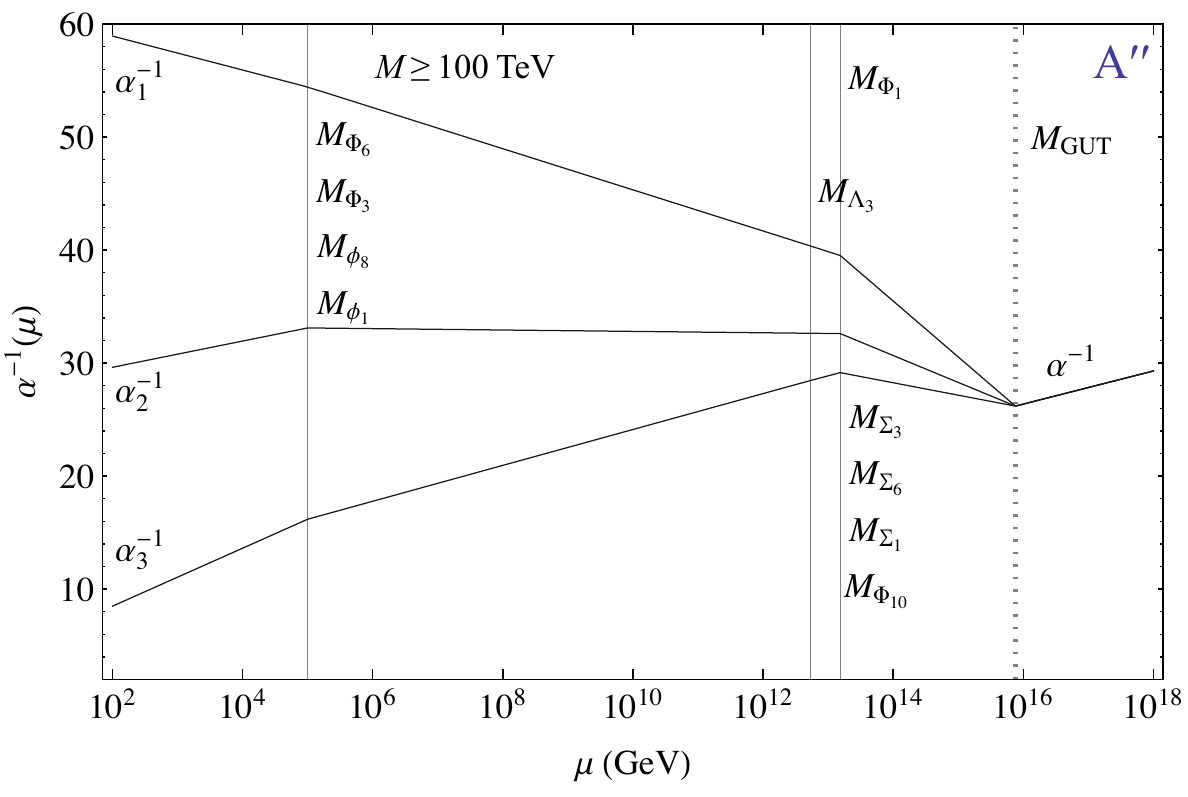}
\caption{Experimentally viable parameter space of the model (left panels) and the gauge coupling unification for the unification points A, A$^\prime$, and A$^{\prime\prime}$ (right panels) when $M \geq 1, 10, 100$\,TeV, as indicated. For additional details see the text.}
\label{fig:MASTER}
\end{figure}

We always set a lower limit on the mass(es) $M_J$ of the new physics state(s), before we numerically look for the viable unification points. We accordingly present our findings for $M \geq 1$\,TeV, $M \geq 10$\,TeV, and $M \geq 100$\,TeV in the first, second, and third row of Fig.~\ref{fig:MASTER}, respectively. In the left three panels of Fig.~\ref{fig:MASTER} we show the contours of constant value of $M_\mathrm{GUT}$ and $\alpha_\mathrm{GUT}$ in $M_{\Phi_1}$-$M_{\Sigma_1}$ plane, where the contours for $M_\mathrm{GUT}$ are given in units of $10^{15}$\,GeV and are shown as the vertical solid lines while the $\alpha_\mathrm{GUT}$ contours are given as dot-dashed lines that run horizontally. We discard the parameter space that corresponds to $M_\mathrm{GUT} \leq 6 \times 10^{15}$\,GeV for the subsequent numerical study in all three instances since our preliminary analysis has shown that such a low $M_\mathrm{GUT}$ is certainly not realistic with regard to the experimental input on the proton decay lifetimes. 

There are two dashed curves in all three panels in the left column of Fig.~\ref{fig:MASTER}. The outermost one represents the boundary after which it is not possible to generate the correct mass scale for neutrinos with perturbative couplings. We generate that curve by setting $\lambda^\prime$ to one and freely varying $M_{\Phi_1}$, $M_{\Sigma_1}$, and six phases in Eq.~\eqref{eq:perturbativity} to find the region where the product $\max(|Y^a_i|) \max(|Y^b_j|)$, $i,j=1,2,3$, does not exceed one. If the product exceeds one we discard that part of parameter space since it can never produce satisfactory neutrino mass fit with perturbative couplings with utmost certainty. The region between the two dashed lines corresponds to the parameter space where it is sometimes possible, for some special choice of the six phases, to find perturbative solution to the neutrino mass fit. Finally, the region to the left of the innermost dashed line yields correct neutrino mass fit for arbitrary choices of the six phases. We also plot the naive bound on the correct neutrino mass scale using green solid contours. These are generated by setting $2 m_0/ \sqrt{\Delta m^2_{31}}$ to 1, 10, and 100, as indicated in the left panels of Fig.~\ref{fig:MASTER}, for $\lambda^\prime=1$. One can see that this naive estimate slightly undershoots the exact result for normal ordering but, still, describes rather accurately the region with the acceptable neutrino mass scale. 

In the three panels in the right column of Fig.~\ref{fig:MASTER} we explicitly present the running of the gauge couplings for one specific unification point, i.e., when $M_{\Phi_1}=M_{\Sigma_1}=10^{13.19}$\,GeV, for the $M \geq 1$\,TeV, $M \geq 10$\,TeV, and $M \geq 100$\,TeV scenarios in the first, second, and third row, respectively, where, again, $M \equiv \min(M_J)$ for $J=\Phi_1, \Phi_3, \Phi_6, \Phi_{10}, \Sigma_1, \Sigma_3, \Sigma_6, \phi_1, \phi_8, \Lambda_3$. The locations of the corresponding unification points in the left panels of Fig.~\ref{fig:MASTER} are denoted with A, A$^\prime$, and A$^{\prime\prime}$.

To clearly demonstrate that $M_\mathrm{GUT}$ does not depend on $M_{\Sigma_1}$ due to the fact that the three multiplets in $15_F$ and $\overline{15}_F$ remain mass degenerate when maximizing $M_\mathrm{GUT}$, we present in Fig.~\ref{fig:OPQ} the running of the gauge couplings for three specific points in the $M_{\Phi_1}$-$M_{\Sigma_1}$ plane corresponding to O$(10^{11.5}\,\mathrm{GeV}, 10^{14.3}\,\mathrm{GeV})$, P$(10^{11.5}\,\mathrm{GeV}, 10^{11.5}\,\mathrm{GeV})$, and Q$(10^{11.5}\,\mathrm{GeV}, 10^{8.7}\,\mathrm{GeV})$, that are clearly marked on the $M \geq 1$\,TeV plot of Fig.~\ref{fig:MASTER}. 

What one can clearly observe from Fig.~\ref{fig:MASTER} is that $\alpha_\mathrm{GUT}$ grows with a decrease in $M_{\Sigma_1}$ while $M_\mathrm{GUT}$ remains constant for fixed $M_{\Phi_1}$. This simply means that the proton decay bound on the model parameter space in the $M_{\Phi_1}$-$M_{\Sigma_1}$ plane is expected to be more stringent as the value of $M_{\Sigma_1}$ decreases. 

The parameter space that we further investigate is shown in the left three panels of Fig.~\ref{fig:MASTER}. It is bounded from the left by the vertical line that corresponds to $M_\mathrm{GUT}=6 \times 10^{15}$\,GeV and from the right by the outermost dashed line after which it is not possible to (re)produce the neutrino mass scale with perturbative couplings. We create a grid of equidistant points within this parameter space in the $M_{\Phi_1}$-$M_{\Sigma_1}$ plane, where the spacing along both axes is 0.1 in units of $\log_{10} (M_{\Phi_1,\Sigma_1}/1\,\mathrm{GeV})$, and then proceed with steps two and three of our numerical analysis of these points as we describe next in more details.

\subsection{Fermion mass fit}
\label{mass_fit}

The Standard Model fermion masses in our model can be read off from 
\begin{equation}
\mathcal{L} \supset -u^TM_Uu^c-d^TM_Dd^c-e^TM_Ee^c-\frac{1}{2} \nu^T M_N \nu+\mathrm{h.c.}\,,
\end{equation}
where $M_U$, $M_D$, $M_E$, and $M_N$ are given in Eqs.~\eqref{eq:massu}, \eqref{eq:massd}, \eqref{eq:masse}, and~~\eqref{eq:nu}, respectively. The fermion mass eigenstate basis, in the most general scenario, is defined through
\begin{align}
&M_U= U_L M_U^\mathrm{diag} U_R^\dagger\,,  \label{eq:massu_a}
\\
&M_D= D_L M_D^\mathrm{diag} D_R^\dagger\,,  \label{eq:massd_a}
\\
&M_E \equiv E_L M_E^\mathrm{diag} E_R^\dagger\,, \label{eq:masse_a}
\\
&M_N=N M_N^\mathrm{diag} N^T\,,
\label{eq:massnu_a}
\end{align}
where $U_L$, $U_R$, $D_L$, $D_R$, $E_L$, $E_R$, and $N$ are the associated unitary transformations. The model stipulates that 
\begin{align}
\label{eq:test_2}
&U_L =D_L \mathrm{diag}(1,e^{i\kappa_4},e^{i\kappa_5}) V^T_\mathrm{CKM} \mathrm{diag}(e^{i \kappa_1},e^{i\kappa_2},e^{i\kappa_3})\,,\\
\label{eq:test_3}
&U_R=U_L^*\,\mathrm{diag}(e^{i \xi_1},e^{i\xi_2},e^{i\xi_3}),\\
\label{eq:test_4}
&E_L=\mathbb{I}\,,\\
\label{eq:test_5}
&E_R=\mathbb{I}\,,\\
\label{eq:test_1}
&N=\mathrm{diag}(e^{i \gamma_1},e^{i\gamma_2},e^{i\gamma_3}) V^*_\mathrm{PMNS}\,,
\end{align}
where $V_\mathrm{CKM}$ is the Cabibbo-Kobayashi-Maskawa (CKM) mixing matrix with one CP violating phase ($\delta^\mathrm{CKM}$) and, again, $V_\mathrm{PMNS}$ is the PMNS mixing matrix with one CP violating phase and two Majorana phases. The connection between $U_L$ and $U_R$ in Eq.~\eqref{eq:test_3} is due to the fact that $M_U=M_U^T$.

To perform the numerical analysis we first take the low-scale experimental values of the charged fermion sector observables and run them up from $M_Z$ to $M_\mathrm{GUT}$ via the relevant one-loop level renormalization group equations~\cite{Arason:1991ic}. During this process, we 
appropriately take into account all the threshold corrections to these observables due to the presence of the new physics states that reside between the low scale and the  unification scale. The mass spectrum of the relevant states is determined by the procedure that is described in Section~\ref{unification}. We repeat the one-loop level renormalization group running for all viable unification points that are presented in the three left panels of Fig.~\ref{fig:MASTER}. (Again, the region of interest is bounded from the left by the vertical line that corresponds to $M_\mathrm{GUT}=6 \times 10^{15}$\,GeV and from the right by the outermost dashed line while the spacing between the neighboring points is 0.1 in units of $\log_{10} (M_{\Phi_1,\Sigma_1}/1\,\mathrm{GeV})$ along both axes.) We then use these evolved quantities associated with a given unification point in our fitting procedure. We present, as an example, the result of the renormalization group running of the $\tau$ $(y_\tau)$, $b$ $(y_b)$, and $t$  $(y_t)$ Yukawa couplings in the right panels of Fig.~\ref{fig:OPQ} for points O, P, and Q, where the new physics mass spectra associated with these unification points are explicitly given in the left panels. Note that the position of the unification points O, P, and Q in the $\Phi_1$-$\Sigma_1$ plane can be read off from the left uppermost panel of Fig.~\ref{fig:MASTER}.
Even though the unification scale is the same for points O, P, and Q, one can observe a 10\,\% fluctuation in the values of aforementioned Yukawa couplings at $M_\mathrm{GUT}$. 

Since the running of the neutrino observables produces very small effect, we fit the corresponding low-scale values. We summarize experimentally measured observables with the associated $1\,\sigma$ uncertainties of both the charged and neutral fermion sectors at low scale in Table~\ref{tab:input}. (We use $v_5(M_Z)=174$\,GeV.) 
\begin{table}[b]
\centering
\footnotesize
\resizebox{0.9\textwidth}{!}{
\begin{tabular}{|c|c|c|c|}
\hline
$m(M_{Z})$ (GeV) & Fit input  & $\theta^{\rm{CKM},\rm{PMNS}}_{ij}$ \& $\delta^{\rm{CKM}}$ \& $\Delta m^2_{ij}$ (eV$^2$) & Fit input  \\
\hline
\hline
$m_{u}/10^{-3}$ & $1.158 \pm 0.392$ &$\sin\theta^{\rm{CKM}}_{12}$ & $0.2254\pm 0.00072$ \\ \hline
$m_{c}$ & $0.627 \pm 0.019$ &$\sin\theta^{\rm{CKM}}_{23}/10^{-2}$ & $4.207\pm 0.064$  \\ \hline
$m_{t}$ & $171.675 \pm 1.506$ &$\sin\theta^{\rm{CKM}}_{13}/10^{-3}$ & $3.640\pm 0.130$  \\ \hline
$m_{d}/10^{-3}$ & $2.864 \pm 0.286$ &$\delta^{\rm{CKM}}$ & $1.208\pm0.054$ \\ \hline
$m_{s}/10^{-3}$ & $54.407 \pm 2.873$ &$\Delta m^{2}_{21}/10^{-5}$ &7.425$\pm$0.205  \\ \hline
$m_{b}$ & $2.854 \pm 0.026$ & $\Delta m^{2}_{3\ell}/10^{-3}$ &2.515$\pm$0.028 \\ \hline
$m_{e}/10^{-3}$ & $0.486576$ &$\sin^{2}\theta^{\rm{PMNS}}_{12}/10^{-1}$  &3.045$\pm$0.125   \\ \hline
$m_{\mu}$ & $0.102719$ &$\sin^{2}\theta^{\rm{PMNS}}_{23}$ &0.554$\pm$0.021  \\ \hline
$m_{\tau}$ & $1.74618$  &$\sin^{2}\theta^{\rm{PMNS}}_{13}/10^{-2}$ &$2.224\pm 0.065$  \\ \hline
\end{tabular}
}
\caption{Experimental observables associated with charged fermions~\cite{Antusch:2013jca} and neutrinos for normal ordering~\cite{Esteban:2020cvm} with 1\,$\sigma$ uncertainties (except for charged leptons).}
\label{tab:input}
\end{table}
We also present in Table~\ref{table:Yukawas} the ranges of values that we find, within the region of interest, for $y_\tau$, $y_b$, and $y_t$ at the unification scale $M_\mathrm{GUT}$ after we implement renormalization group running procedure of the central values of these quantities as given in Table~\ref{tab:input}. 
\begin{table}
\begin{center}
\begin{tabular}{| l | c | c | c |}
\hline
$M \equiv \min(M_J)$ & $y_\tau(M_\mathrm{GUT})/10^{-3}$ & $y_b(M_\mathrm{GUT})/10^{-3}$ & $y_t(M_\mathrm{GUT})$ \\
\hline
\hline
$M \geq 1$\,TeV & $(8.36,9.05)$ & $(3.68,4.70)$ & $(0.263,0.342)$\\
$M \geq 10$\,TeV & $(8.70,9.27)$ & $(4.23,5.08)$ & $(0.306,0.372)$\\
$M \geq 100$\,TeV & $(9.20,9.43)$ & $(5.04,5.36)$ & $(0.371,0.396)$\\
\hline  
\end{tabular}
\end{center}
\caption{The ranges of values of Yukawa couplings of the third generation Standard Model fermions after the one-loop running of the central values from $M_Z$ to $M_\mathrm{GUT}$ for the unification points that reside within the regions of interest shown in the left three panels of Fig.~\ref{fig:MASTER}.}
\label{table:Yukawas}
\end{table}

In the numerical fit we use charged lepton masses at $M_\mathrm{GUT}$ as an input to determine $Y^d$ in a straightforward fashion since $v_5 Y^d= M_E^\mathrm{diag}= \mathrm{diag}(m_e,m_\mu,m_\tau)$. The model thus addresses charged lepton masses exactly.
The down-type quark mass matrix of Eq.~\eqref{eq:massd} and the neutrino mass matrix of Eq.~\eqref{eq:massnu} share a common Yukawa coupling row matrix $Y^a$. We accordingly perform a combined fit to data for these two sectors. 
To that end, we minimize a $\chi^2$ function which is defined as
\begin{align}
\chi^2=\sum_k P^2_k, \;\;\; P_k= \frac{T_k-O_k}{E_k}\,,    
\end{align}
where, $T_k$, $O_k$, and $E_k$ represent theoretical prediction, measured central value, and experimental $1\,\sigma$ error for the observable $k$, respectively. $k$ runs over the neutrino sector observables and the down-type quark masses. Clearly, in our fitting approach, the Yukawa coupling matrices $Y^{a}$, $Y^{b}$, and $Y^{c}$ are determined against three down-type quark masses, two neutrino mass-squared differences and three mixing angles in the neutrino sector. (The CP violating phase and the two Majorana phases in the neutrino sector have not been experimentally measured.) We scan over all viable unification points demanding perturbativity of the relevant couplings, i.e., $\max(|Y^a_i|), \max(|Y^b_i|), \max(|Y^c_i|), |\lambda^\prime | \leq 1$, and utilize the criteria $\chi^2/n \leq 1$ to be considered as a good fit, where $n(=8)$ is the number of fitted observables. (Note that not all the unification points that allow for a good numerical fit pass the proton decay test, to be detailed in Section~\ref{proton_decay}.) 

We point out that our combined numerical fit of the down-type quark and neutrino sectors demonstrates that this model cannot accommodate the inverted neutrino mass ordering.
Note that $Y^d$ is a hierarchical diagonal matrix, where its entries are completely determined by the charged lepton Yukawa couplings. Since the matrix elements $(M_D)_{ij}$ are proportional to the linear combination of $(Y^d)_{ij}$ and $Y^c_i Y^a_j$ it is obvious that $Y^a$ and $Y^c$ should both be hierarchical row matrices to produce a good fit to data. This, however, is impossible to achieve for the inverted ordering of the neutrino masses. Namely, for the inverted scenario the entries in the first row and the first column of the neutrino mass matrix $M_N$ are typically of the same order, whereas the lower $2\times 2$ block is required to be somewhat smaller in magnitude. (Note that we work in the mass eigenstate basis for the charged leptons.) This requirement forces all entries in $Y^a$ to be of the same order which is in direct conflict with what is needed in the down-type quark sector. This tension is a direct consequence of the simplicity of the model which, in turn, leads to a prediction that neutrinos must have the normal mass ordering.

The fitting procedure, for the normal ordering of neutrino masses, allows us to numerically determine three unitary rotation matrices $D_L$, $D_R$, and $N$, as well as the Yukawa couplings of the charged leptons and the down-type quarks. To fully compute partial lifetimes for different proton decay modes, due to both the gauge boson and scalar mediations, one also needs to know the unitary matrices $U_L$ and $U_R$ that diagonalize the up-type quark mass matrix and the associated Yukawa couplings. The nice feature of our approach is that the former can be expressed in terms of $D_L$ and $V_{CKM}$, as given in Eqs.~\eqref{eq:test_2} and~\eqref{eq:test_3}, and eight additional phases. The latter can also be found since 
\begin{equation}
M_U=4 v_5(Y^u+Y^{u\,T})=U_L \mathrm{diag}(m_u,m_c,m_t )U^\dagger_R\,,
\end{equation}
where $U_L$ and $U_R$, again, are connected via Eq.~\eqref{eq:test_3}.

To summarize, the model accommodates charged lepton masses, the up-type quark masses, and the CKM parameters exactly. We furthermore perform combined numerical fit of the neutrino mass parameters, the down-type quark masses, and the PMNS parameters since these are inextricably linked. The most important outcome of the fit for the proton decay considerations are the unitary transformations $U_L$, $U_R$, $D_L$, and $D_R$, where the first two matrices feature five and three unknown phases, respectively. We will show next that the analysis of the leading source of proton decay requires knowledge of only two of these phases, i.e., $\kappa_4$ and $\kappa_5$ of Eq.~\eqref{eq:test_2}, that reside in $U_L$.

\subsection{Proton decay analysis}
\label{proton_decay}

The main constraint on the otherwise viable parameter space of the model originates from the experimental limit on the partial lifetime of the $p \rightarrow \pi^0 e^+$ process. Since the maximal possible value of $M_\mathrm{GUT}$ and the associated mass spectrum of the new physics states are known for every unification point, with $M_\mathrm{GUT}$ being the mass of the proton decay mediating gauge bosons in $24_V$, we can set an accurate lower bound on $M_\mathrm{GUT}$ due to the proton decay lifetime measurements through the use of the numerical output of the fermion sector fit from the previous section. The result of our study is presented in the left panels of Fig.~\ref{fig:MASTER}, where the region to the left of the boundary that is marked with ``proton decay bound'' wording, in the $\Phi_1$-$\Sigma_1$ plane, is already excluded with the current  data. We stress that we have looked at all two-body proton decay signatures to find that it is the $p \rightarrow \pi^0 e^+$ partial lifetime limit that is the most constraining. We present, in what follows, the procedure that we use to produce the proton decay bounds of the left panels of Fig.~\ref{fig:MASTER}. 

The relevant proton decay width for the $p \to \pi^0 e^+$ process is~\cite{Nath:2006ut}
\begin{align}
\nonumber
\Gamma (p \to \pi^0 e^+)&= \frac{m_p \pi}{2} \left( 1-\frac{m_\pi^2}{m_p^2} \right)^2 A_L^2 \frac{\alpha_\mathrm{GUT}^2}{M_\mathrm{GUT}^4} \\
\nonumber
&\times \left(  A^2_{SL}\left| c(e^c,d) \matrixel{\pi^0}{(ud)_L u_L}{p}\right |^2  + A^2_{SR} \left| c(e,d^c) \matrixel{\pi^0}
{(ud)_R u_L}{p}\right|^2 \right)\,, 
\end{align}
where the relevant matrix elements are $\matrixel{\pi^0}{(ud)_Lu_L}{p}=0.134(5)(16)\,\mathrm{GeV}^2$ and  $\matrixel{\pi^0}{(ud)_Ru_L}{p}=-0.131(4)(13)\,\mathrm{GeV}^2$~\cite{Aoki:2017puj}, $A_L(=1.2)$ captures the QCD running of the proton decay operators below the $M_Z$ scale~\cite{Nihei:1994tx}, $m_p(=0.9393$\,GeV) is the proton mass, and $m_\pi(=0.134$\,GeV) is the pion mass. The running of the proton decay operators from the unification scale $M_\mathrm{GUT}$ down to $M_Z$ is given by $A_{SL}$ and $A_{SR}$, where these coefficients
are~\cite{Buras:1977yy,Ellis:1979hy,Wilczek:1979hc}
\begin{equation*}
A_{SL(R)}=\prod_{i=1,2,3} \prod_I^{M_Z \leq M_I \leq M_{GUT}}
\left[\frac{\alpha_i(M_{I+1})}{\alpha_i(M_I)}\right]^{\frac{\gamma_{L(R)i}}{\sum_J^{M_Z
\leq M_J \leq M_I} b^J_{i}}},\,
\gamma_{L(R)i}=(23(11)/20,9/4,2).
\end{equation*}
Indices $I$ and $J$ run through all the new physics states that reside below the unification scale. We evaluate $A_{SL}$ and $A_{SR}$ for every point that provides satisfactory unification as well as viable fit of the fermion observables using the associated mass spectrum. 

The coefficients $c(e^c, d)$ and $c(e, d^c)$, in our model, are
\begin{align}
\label{eq:cL}
&c(e^c_\alpha, d_\beta) = e^{-i \xi_1} \left( (D_L^*)_{11} + ( U_L^T D_L^*)_{1 1} (U_L^*)_{1 1} \right),\\
\label{eq:cR}
&c(e, d^c) = e^{-i \xi_1} (D_R^\dagger)_{11}\,.
\end{align}
One can observe that the phase $\xi_1$ of Eq.~\eqref{eq:test_3} does not enter the predictions for the $p \to \pi^0 e^+$ decay width. Moreover, out of the five phases in $U_L$, as given in Eq.~\eqref{eq:test_2}, only $\kappa_4$ and $\kappa_5$ affect the value of $c(e^c_\alpha, d_\beta)$ in Eq.~\eqref{eq:cL}. Since the matrix elements of $D_L$ and $D_R$ are obtained during the numerical fitting procedure and $U_L$ can be expressed in terms of the CKM matrix via Eq.~\eqref{eq:test_2}, all we need to do is to vary $\kappa_4$ and $\kappa_5$ to find the smallest possible value for $|c(e^c_\alpha, d_\beta)|$ and hence the most conservative bound on the parameter space of the model due to the experimental limit on the partial proton lifetime for the $p\to \pi^0 e^+$ channel. To produce the bounds in the left panels of Fig.~\ref{fig:MASTER} we use $\tau^\mathrm{exp}_{p\to \pi^0 e^+} > 2.4 \times 10^{34}$\,years, as given by the Super-Kamiokande Collaboration~ \cite{Takenaka:2020vqy}. (The Hyper-Kamiokande~\cite{Abe:2018uyc} detector has the potential to significantly reduce the allowed parameter space of our model. The expected $90\,\%$ confidence level on the $p\to \pi^0e^+$ channel for 10 years (20 years) of operation is $7.8\times 10^{34}$ ($1.3\times 10^{35}$) years, whereas the $3\,\sigma$ discovery potential reach is $6.3\times 10^{34}$ ($1.0\times 10^{35}$) years.)

Note that the proton decay bound in the left panels of Fig.~\ref{fig:MASTER} slopes slightly to the right as the mass of $\Sigma_1$ decreases. The main reason for that is the fact that  $\alpha_\mathrm{GUT}$ grows with a decrease in the $\Sigma_1$ mass for a fixed value of $M_{\Phi_1}$ whereas $M_\mathrm{GUT}$ remains constant. Also, we can predict the extent to which one needs to experimentally improve the limit on the $p\to \pi^0 e^+$ partial lifetime in order to completely rule out currently available parameter space. An improvement of the current $p \rightarrow \pi^0 e^+$ lifetime limit by a factor of $2$, $15$, and $96$ would completely rule out the $M \geq 100$\,TeV, $M \geq 10$\,TeV, and $M \geq 1$\,TeV scenarios, respectively. The last point to be eliminated in the $\Phi_1$-$\Sigma_1$ plane, in all three left panels of Fig.~\ref{fig:MASTER}, by such an improvement is $(M_{\Phi_1}=10^{13.2}\,\mathrm{GeV}, M_{\Sigma_1}=10^{13.6}\,\mathrm{GeV})$. In fact, the scenario where the lower bound on the masses of the new physics states is set at $120$\,TeV is already completely ruled out by the proton decay constraints. The states that prefer to be light, in order to maximize the unification scale, are scalars $\phi_1$, $\phi_8$, $\Phi_3$, and $\Phi_6$.     

\subsection{Results}
\label{results}

In this section we succinctly summarize our numerical findings.

The viable parameter space of the model is given in the three left panels of Fig.~\ref{fig:MASTER} in the $M_{\Phi_1}$-$M_{\Sigma_1}$ plane, where we show the contours of constant $M_\mathrm{GUT}$, $\alpha_\mathrm{GUT}$, and $m_0$ for $|\lambda^\prime|=1$. The unification scale contours are given in units of $10^{15}$\,GeV and appear as vertical solid lines while the $\alpha_\mathrm{GUT}$ contours are given as dot-dashed lines that run horizontally. The contours of constant $m_0$, for $|\lambda^\prime|=1$, are shown as green solid curves. The unification scale is maximized by freely varying masses $M_J$, where $J=\Phi_1, \Phi_3, \Phi_6, \Phi_{10}, \Sigma_1, \Sigma_3, \Sigma_6, \phi_1, \phi_8, \Lambda_3$, while taking into account additional constraints discussed in Section~\ref{GCU} and imposing a condition that $M \equiv \min(M_J)$ is greater or equal to $1$\,TeV, $10$\,TeV, and $100$\,TeV in the panels of the first, second, and third row of the left column of Fig.~\ref{fig:MASTER}, respectively.

In the right panels of Fig.~\ref{fig:MASTER} we present the running of the gauge couplings for one particular unification point, i.e., when $M_{\Phi_1}=M_{\Sigma_1}=10^{13.19}$\,GeV, for the $M \geq 1$\,TeV, $M \geq 10$\,TeV, and $M \geq 100$\,TeV scenarios in the first, second, and third row, respectively. The unification points that correspond to these new physics mass spectra are denoted with A, A$^\prime$, and A$^{\prime\prime}$ in the left panels of Fig.~\ref{fig:MASTER}.

The parameter space that is viable with respect to the experimental input, for the three cases at hand, can be read off from the left panels of Fig.~\ref{fig:MASTER}. Namely, it is bounded from the left by the proton decay bound, as discussed in Section~\ref{proton_decay}, and from the right by the outermost dashed curve. The outermost dashed curve delineates the region after which it is not possible to address phenomenologically viable neutrino mass scale with perturbative couplings.

The proton decay bound in Fig.~\ref{fig:MASTER} is generated by the current experimental limit on the partial lifetime for the $p \rightarrow \pi^0 e^+$ process. We find that an improvement of the current $p \rightarrow \pi^0 e^+$ lifetime limit by a factor of $2$, $15$, and $96$ would completely rule out the $M \geq 100$\,TeV, $M \geq 10$\,TeV, and $M \geq 1$\,TeV scenarios, respectively. The last viable point to be eliminated by the aforementioned improvement, in all three left panels of Fig.~\ref{fig:MASTER}, is $(M_{\Phi_1},M_{\Sigma_1})=(10^{13.2}\,\mathrm{GeV}, 10^{13.6}\,\mathrm{GeV})$. This is to be expected since  $\alpha_\mathrm{GUT}$ grows with a decrease in the $\Sigma_1$ mass for a fixed value of $M_{\Phi_1}$, whereas $M_\mathrm{GUT}$ remains constant. To demonstrate the latter we present unification for points O, P, and Q in the left panels of Fig.~\ref{fig:OPQ}, whereas the associated renormalization group running of the $\tau$ $(y_\tau)$, $b$ $(y_b)$, and $t$  $(y_t)$ Yukawa couplings is shown in the right panels.

Our numerical fit explicitly yields all unitary transformations and Yukawa couplings except for eight phases associated with the up-type quarks sector. As we do not use the CP phase in the neutrino sector as an input for our numerical fit we find, within the viable parameter space that is shown in the left panels of Fig.~\ref{fig:MASTER}, that $\delta^\mathrm{PMNS} \in
\left(-35.6^{\circ},+29.9^{\circ}\right)$ for $M \geq 1\,\mathrm{TeV}$, $\delta^\mathrm{PMNS} \in
\left(-43.06^{\circ},+40.18^{\circ}\right)$ for $M \geq 10\,\mathrm{TeV}$, and $\delta^\mathrm{PMNS} \in
\left(-47.6^{\circ},+53.0^{\circ}\right)$ for $M \geq 100\,\mathrm{TeV}$.

\section{Conclusion}
\label{SEC-04}

We present phenomenological study of the viable parameter space of the most minimal realistic $SU(5)$ model. The structure of the model is built entirely out of the fields residing in the first five lowest lying representations in terms of dimensionality that transform non-trivially under the $SU(5)$ gauge group. The model has eighteen real parameters and fourteen phases to address experimental observables of the Standard Model fermions and accomplishes that via simultaneous use of three different mass generation mechanisms. It inextricably links the origin of the neutrino mass to the experimentally observed difference between the down-type quark and charged lepton masses. The main predictions of the model are that $(i)$ the neutrinos are Majorana particles, $(ii)$ one neutrino is massless, $(iii)$ the neutrinos have normal mass ordering, and $(iv)$ there are four new scalar multiplets at or below a $120$\,TeV mass scale. An improvement of the current $p \rightarrow \pi^0 e^+$ lifetime limit by a factor of $2$, $15$, and $96$ would require these four scalar multiplets to reside at or below the $100$\,TeV, $10$\,TeV, and $1$\,TeV mass scales, respectively. The numerical analysis of the model also yields a range of viable values for $\delta^\mathrm{PMNS}$, i.e., the CP phase in the PMNS matrix, as a function of the lower limit $M \equiv \min(M_J)$ on the masses $M_J$ of the new physics states $J$, where $J=\Phi_1, \Phi_3, \Phi_6, \Phi_{10}, \Sigma_1, \Sigma_3, \Sigma_6, \phi_1, \phi_8, \Lambda_3$. These ranges are $\delta^\mathrm{PMNS} \in
\left(-35.6^{\circ},+29.9^{\circ}\right)$ for $M \geq 1\,\mathrm{TeV}$, $\delta^\mathrm{PMNS} \in
\left(-43.06^{\circ},+40.18^{\circ}\right)$ for $M \geq 10\,\mathrm{TeV}$, and $\delta^\mathrm{PMNS} \in
\left(-47.6^{\circ},+53.0^{\circ}\right)$ for $M \geq 100\,\mathrm{TeV}$.

\FloatBarrier

\FloatBarrier

\section*{Acknowledgments}
I.D.\ would like to thank the CERN Theory Department on hospitality and support through the Corresponding Associates program. The work of S.S.\ has been supported by the Swiss National Science Foundation.

\bibliographystyle{style}
\bibliography{references}
\end{document}